\documentclass[12pt]{iopart}

\usepackage{iopams,color}

\def\be{\begin{equation}}
\def\ee{\end{equation}}
\def\ba{\begin{eqnarray}}
\def\ea{\end{eqnarray}}

\newtheorem{definition}{Definition}[section]

\newtheorem{prop}[definition]{Proposition}

\begin{document}
	
\begin{flushright}
	KOBE-COSMO-16-05
\end{flushright}
	
\title[]{Rational first integrals of geodesic equations
	and generalised hidden symmetries
}
	
\author{Arata Aoki, Tsuyoshi Houri and Kentaro Tomoda
}
	
\address{Department of Physics, Kobe University, 1-1 Rokkodai, Nada, Kobe, Hyogo, 657-8501 JAPAN
}

\eads{\mailto{arata.aoki@stu.kobe-u.ac.jp}, \mailto{houri@phys.sci.kobe-u.ac.jp}, 
	\mailto{k-tomoda@stu.kobe-u.ac.jp}
}

\begin{abstract}
We discuss novel generalisations of Killing tensors, which are introduced by considering rational first integrals of geodesic equations.
We introduce the notion of inconstructible generalised Killing tensors,
which cannot be constructed from ordinary Killing tensors.
Moreover, we introduce inconstructible
rational first integrals, which are constructed from inconstructible generalised Killing tensors, and provide a method for checking the inconstructibility of a rational first integral.
Using the method, we show that the rational first integral of the Collinson-O'Donnell solution is not inconstructible.
We also provide several examples of metrics admitting an inconstructible rational first integral in two and four dimensions, by using the Maciejewski-Przybylska system.
Furthermore, we attempt to generalise other hidden symmetries such as Killing-Yano tensors.
\end{abstract}

	
\maketitle

\section{Introduction}
In general relativity, hidden symmetries of spacetime such as
Killing tensors \cite{Stackel:1895} and
Killing-Yano tensors \cite{Yano:1952,Tachibana:1969,Kashiwada:1968,Kora:1980}
have helped us in understanding various phenomena
in a strong gravitational field.
An important example is the Kerr spacetime,
which describes an isolated stationary rotating black hole in a vacuum.
In the Kerr spacetime, the geodesic equations can be solved
by separation of variables due to the presence of a Killing tensor \cite{Carter:1968}.
It is also known that the Klein-Gordon and Dirac equations can be solved
by separation of variables due to the presence of a Killing-Yano tensor
\cite{Carter:1977,Carter:1979}.
Moreover, if a four-dimensional spacetime possesses a (nondegenerate rank-2) 
Killing-Yano tensor,
the canonical form for metrics is provided in the Carter-Plebanski form \cite{Dietz:1982}
and it is shown that the Kerr spacetime with a NUT parameter
is the only vacuum solution of the Einstein equations
with such hidden symmetry \cite{Carter:1968,Plebanski:1975}.
In recent years, hidden symmetries of higher-dimensional black hole
spacetimes have been uncovered
(see, e.g., \cite{Frolov:2008,Yasui:2011} for reviews).
In mathematics, (conformal) Killing tensors \cite{Heil:2015}
and (conformal) Killing-Yano tensors \cite{Semmelmann:2002}
have been studied from the modern geometric point of view.

The purpose of this paper is to discuss novel generalisations of
Killing tensors, as well as Killing-Yano tensors,
to stretch the concept of hidden symmetries of spacetime.
Various generalisations have been proposed in the past.
A remarkable one is the generalised (conformal) Killing-Yano tensors
introduced by Wu \cite{Wu:2009} and
Kubiz\v{n}\'ak, Kunduri and Yasui \cite{Kubiznak:2009}.
The authors replaced the Levi-Civita connection in the Killing-Yano
equation by connections with totally skew-symmetric torsion
\footnote{It should be remarked that the idea of considering connections
	with totally skew-symmetric torsion was already 
	proposed by Strominger \cite{Strominger:1986}
	in the context of string theories, where such a torsion is identified
	with the 3-form flux livinging in the theories.}.
Recently, the generalised (conformal) Killing-Yano tensors
have been studied well in the context of string theories 
\cite{Houri:2010,Houri:2013,Chow:2015,Chervonyi:2015,Santillan:2012,Batista:2015}.
On the other hand, such a generalisation does not affect the Killing equation.
Killing equations defined by connections with totally skew-symmetric torsion are,
if the connections satisfy the metric condition, completely equivalent
to the ordinary Killing equation with the Levi-Civita connection.

To generalise Killing tensors, we consider certain connections
which are torsion-free but do not satisfy the metric condition.
We replace the Levi-Civita connection in the Killing equation
by these connections.
Physically, this generalisation is justified
by considering rational first integrals of geodesic equations.
It is known that ordinary Killing tensors are introduced by
considering polynomial first integrals of geodesic equations.
The condition that a polynomial function is a first integral
leads us to the Killing equation.
In analogy with this, when we consider the condition
that a rational function is a first integral,
we are naturally led to introduce the generalised Killing equation
(see Sec.\ II for details).

In this paper, we shall restrict our analysis to the geodesic equations
on a space or spacetime $(M,g_{\mu\nu})$.
The Hamiltonian
is given by
\be
H = \frac{1}{2} g^{\mu\nu}p_\mu p_\nu \,, \label{Hamiltonian}
\ee
where $g^{\mu\nu}$ is the inverse of $g_{\mu\nu}$.
We will use $x^\mu$ and $p_\mu$ as canonical coordinates
and momenta of a particle on $M$, respectively.
Geometrically, $(x^\mu,p_\mu)$ may be considered as local coordinates on $T^*M$.
Following the Kozlov's notaion \cite{Kozlov:2014}, we consider a rational first integral $F$,
which has the form
\be
F = \frac{P}{Q} \,, \label{basic1}
\ee
where $P$ and $Q$ are nonzero polynomials of degree $r$ and $s$ in $p_\mu$,
i.e., $F$ is rational in $p_\mu$.
We assume that $r \geq s$, which can be done without loss of generality
because, if $F$ is a first integral, $F^{-1}$ is also a first integral.
We also assume that $P$ and $Q$ are relatively prime
in the sense that 
	as polynomials in $p_\mu$ they have no common root.
Moreover, a rational first integral is said to be {\it irreducible}
if the degrees of $P$ and $Q$ cannot be reduced by using the Hamiltonian
or other first integrals.
For example, if $p_y/p_x$ is a first integral,
$(p_y+Hp_x)/p_x$ and $(p_x^2+p_y^2)/p_xp_y$ are also first integrals
but they are not irreducible.

Besides, we introduce the notion of {\it inconstructible}
rational first integrals, which has been little discussed
in previous works.
It is obvious that if $P$ and $Q$ are first integrals,
$F$ is a first integral.
Still obvious is that if there exists
a function on $M$, $\psi=\psi(x^\mu)$,
such that $\bar{P}=\psi P$ and $\bar{Q}=\psi Q$
are first integrals, $F=\bar{Q}/\bar{P}$ is a first integral.
This suggests that the rational first integrals which can be
constructed from two polynomial first integrals are not so meaningful.
Hence, we distinguish the ones from the others
which cannot be constructed from two first integrals.
Such rational first integrals are said to be inconstructible.
Although there is a possibility that $\psi$ is a function on $T^*M$,
$\psi=\psi(x^\mu,p_\mu)$, we do not think about such a case in this paper.

In mathematical physics, the study of rational first integrals
was already initiated by Darboux \cite{Darboux:1896}.
Related to it, a lot of works have been conducted.
Yet, many of them are about autonomous systems
and their significance in general relativity is unclear.
A striking example in general relativity is the Collinson-O'Donnell 
solution \cite{Collinson:1992},
which is a solution of the vacuum Einstein equations
admitting a rational first integral of the geodesic equations.
However, as we will show in Sec.\ III, the rational first integral
is not inconstructible, so that the Collinson-O'Donnell 
solution is not a nontrivial example in that sense.
Hence, it is another purpose of this paper to obtain nontrivial examples
for solutions in general relativity admitting an inconstructible rational first integral.

This paper is organised as follows: In the next section, we discuss the conditions
that $F$ is a first integral of geodesic equations
and show that they naturally introduce a generalisation of Killing tensors.
For Killing vectors, this was introduced by Collinson \cite{Collinson:1986}.
After introducing the notion of inconstructible generalised Killing tensors,
we provide a method for checking whether a generalised Killing tensor is
inconstructible.
We also show that the defining equation of the generalised Killing tensors
can be written in the same form as ordinary Killing tensors, where
the Levi-Civita connection is replaced by certain connections.
Furthermore, we provide the integrability conditions
for generalised Killing vectors in terms of the present connections.
In Sec.\ III, by using the method we provided in Sec.\ II,
we investigate whether the rational first integral
of the Collinson-O'Donnell solution is inconstructible.
In Sec.\ IV, we construct several metrics admitting an inconstructible rational first integral
in two and four dimensions, by using the Maciejewski-Przybylska system \cite{Maciejewski:2004}.
At the same time, we investigate their geometric properties
described by the metrics obtained.
In Sec.\ V, we generalise other hidden symmetries:
conformal Killing tensors, Killing-Yano tensors
and conformal Killing-Yano tensors.
Sec.\ VI is devoted to summary and discussion.

\section{Generalised Killing tensors}

\subsection{Rational first integrals and generalised Killing tensors}
The condition that $F$, given by Eq.\ \eref{basic1},
is a first integral for a Hamiltonian $H$, given by Eq.\ \eref{Hamiltonian},
leads to 
\be
 \{P,H\}Q - \{Q,H\}P = 0 \,, \label{eq3}
\ee
where $\{~,~\}$ is the Poisson bracket.
Introduced an auxiliary function $A$,
it is equivalent to
\be
 \{P,H\} = AP \,, \quad \{Q,H\} = AQ \,, \label{basic3}
\ee
where $A$ is called a {\it cofactor} of $P$ and $Q$.

	When $P$ and $Q$ are homogeneous polynomials, they are written as $P = \xi^{\mu_1\cdots \mu_r}p_{\mu_1}\cdots p_{\mu_r}$ and $Q = \eta^{\mu_1\cdots \mu_s}p_{\mu_1}\cdots p_{\mu_s}$, where $\xi_{\mu_1\cdots\mu_r}$ and
 $\eta_{\mu_1\cdots\mu_s}$ are totally symmetric tensors.
	Substituting these expressions into Eqs.\ \eref{basic3} together with the Hamiltonian (\ref{Hamiltonian}) and evaluating it order by order in $p_\mu$, we find that $A$ is required to be a polynomial of linear order in $p_\mu$, so that writing $A$ as $A = f^\mu p_\mu$, we are able to rewrite Eqs.\ \eref{basic3} as
\be
 \nabla_{(\nu}\xi_{\mu_1\cdots \mu_r)} = f_{(\nu}\xi_{\mu_1\cdots \mu_r)} \,, \quad
 \nabla_{(\nu}\eta_{\mu_1\cdots \mu_s)} = f_{(\nu}\eta_{\mu_1\cdots \mu_s)} \,, \label{GKSeq}
\ee
where $\nabla_\mu$ denotes the Levi-Civita connection and
the round bracket denotes symmetrisation of indices.
If $P$ and $Q$ are inhomogeneous polynomials, we divide them into the parts by order, i.e., $P=\sum_{k=0}^{r}P_k$ and $Q=\sum_{k=0}^{s}Q_k$ where $P_k$ and $Q_k$ denote the order-$k$ parts of $P$ and $Q$, respectively. We then find in the similar fashion above that since the Hamiltonian (\ref{Hamiltonian}) is homogeneous, $A$ is required to be a polynomial of linear order in $p_\mu$, and $P_k$ and $Q_k$ satisfy
\be
\{P_k,H\} = AP_k \,, \quad \{Q_k,H\} = AQ_k \,, \label{basic3_2}
\ee
for every $k$, which lead to Eqs.\ (\ref{GKSeq}) order by order in $p_\mu$. Hence, if a Hamiltonian is homogeneous, we have only to consider the case when $P$ and $Q$ are homogeneous. Eqs.\ (\ref{GKSeq}) have the same form as the Killing tensor equations if $f_\mu=0$. This motivates us to introduce the following definition for generalised Killing tensors.

\begin{definition}
A symmetric tensor $K_{\mu_1\cdots \mu_p}$ is called a generalised Killing tensor
if there exists a 1-form $f_\mu$ satisfying the differential equation
\be
\nabla_{(\nu}K_{\mu_1\cdots \mu_p)}
= f_{(\nu}K_{\mu_1\cdots \mu_p)} \,, \label{GKSeq2}
\ee
where $f_\mu$ is called the associated 1-form of $K_{\mu_1\cdots \mu_p}$.
In particular, a generalised Killing tensor is called a Killing tensor
if the associated 1-form vanishes.
\end{definition}

For rank 1, they are the generalised Killing vectors
introduced by Collinson \cite{Collinson:1986}.
To obtain a rational first integral of geodesic equations,
we need to find a pair of generalised Killing tensors $\xi_{\mu_1\cdots\mu_r}$ and $\eta_{\mu_1\cdots\mu_s}$
satisfying the conditions \eref{GKSeq}
with a common associated 1-form $f_\mu$.
This pair is called a Killing pair \cite{Vaz:1992,Collinson:1992}.

It is worth commenting that in the defining equation (\ref{GKSeq2}), the associated 1-form $f_\mu$ is uniquely determined. If there were two different associated 1-forms $f^{(1)}_\mu$ and $f^{(2)}_\mu$, it would lead to the existence of the 1-form $k_\mu\equiv f^{(2)}_\mu-f^{(1)}_\mu$ which satisfies
\be
 k_{(\nu}K_{\mu_1\cdots\mu_p)}=0 \,. \label{forproof}
\ee 
However, such a 1-form does not exist because it is possible to show that if one component of $k_\mu$ is nonzero, all the components of $K_{\mu_1\cdots\mu_p}$ must vanish: Let $k_i$ be a nonzero component of $k_\mu$ $(0 \leq i \leq n= \dim M)$. Then the $(i, i, \cdots, i)$ component of Eq.\ \eref{forproof}, $k_i K_{i i \cdots i} = 0$, induces $K_{i i \cdots i} = 0$. Next, for $j \neq i$, the $(i, i, \cdots, i, j)$ component of Eq.\ \eref{forproof}, $n K_{ii\cdots ij} k_i + K_{ii\cdots ii} k_j =0$, leads to $K_{ii\cdots ij} = 0$. Moreover, for $\ell \neq j, i$, the $(i, i, \cdots, i, j, \ell)$ component of Eq.\ \eref{forproof}, $(n-1) K_{ii\cdots ij\ell} k_i + K_{ii\cdots ij} k_\ell + K_{ii \cdots i\ell} k_j =0$, leads to $K_{ii\cdots ij\ell} = 0$. In the repetitive manner, it is shown that all the components of $K_{\mu_1\cdots\mu_p}$ vanish.

We find some properties of the generalised Killing tensors.
Given two generalised Killing tensors,
their symmetric tensor product is also a generalised Killing tensor.
If two generalised Killing tensors have the common associated 1-form,
their linear combination is also a generalised Killing tensor.
The following property is about
a functional multiplication of a generalised Killing tensor.

\begin{prop}
Suppose $K_{\mu_1\cdots\mu_p}$ is a generalised Killing tensor.
Then, $\bar{K}_{\mu_1\cdots\mu_p}\equiv \psi K_{\mu_1\cdots\mu_p}$
is also a generalised Killing tensor for an arbitrary function $\psi$.
\end{prop}
{\it Proof.}\quad
Since $K_{\mu_1\cdots\mu_p}$ satisfies Eq.\ \eref{GKSeq2}, we have
\ba
\nabla_{(\nu}\bar{K}_{\mu_1\cdots\mu_p)}
&=& K_{(\mu_1\cdots\mu_p}\partial_{\nu)}\psi 
+ f_{(\nu}K_{\mu_1\cdots \mu_p)} \psi \nonumber\\
&=& \bar{f}_{(\nu}\bar{K}_{\mu_1\cdots \mu_p)}\,,
\ea
where $\bar{f}_\mu=f_\mu + \partial_\mu \ln \psi$.\hfill$\Box$\\

From this proposition, it turns out that a functional
multiplication of a Killing tensor
is also a generalised Killing tensor.
However, not all generalised Killing tensors
can be written as a functional multiplication of a Killing tensor.
A generalised Killing tensor $K_{\mu_1\cdots\mu_p}$
is said to be {\it inconstructible} if there exists no function $\psi$
such that $\bar{K}_{\mu_1\cdots\mu_p}\equiv \psi K_{\mu_1\cdots\mu_p}$
is a Killing tensor.
This notion is important due to the fact that
if we construct a rational first integral from
two constructible generalised Killing tensors with a common associated 1-form, 
the first integral obtained becomes constructible.

\begin{prop}
A generalised Killing tensor is constructible if
and only if the associated 1-form is closed.
\end{prop}
{\it Proof.} Let $K_{\mu_1\cdots\mu_p}$ be a generalised Killing tensor, which satisfies Eq.\ \eref{GKSeq2}. (if) Since the associated 1-form $f_\mu$ is closed, $\nabla_{[\mu}f_{\nu]}=0$,
there exists a function $\psi$
such that $f_\mu=\partial_\mu \ln \psi$.
Using this function, we define
$\bar{K}_{\mu_1\cdots\mu_p}\equiv \psi^{-1} K_{\mu_1\cdots\mu_p}$ and
 find that $\bar{K}_{\mu_1\cdots\mu_p}$ is a Killing tensor,
which satisfies the Killing equation $\nabla_{(\nu}\bar{K}_{\mu_1\cdots\mu_p)}=0$.
(only if) Since $K_{\mu_1\cdots\mu_p}$ is constructible, there exists a function $\psi$ such that $\bar{K}_{\mu_1\cdots\mu_p}\equiv \psi^{-1} K_{\mu_1\cdots\mu_p}$ is a Killing tensor.
Using this, we obtain
\be
 \nabla_{(\nu}K_{\mu_1\cdots\mu_p)}
= \nabla_{(\nu}\Big(\psi\bar{K}_{\mu_1\cdots\mu_p)}\Big)
= \Big(\partial_{(\nu}\ln \psi\Big) K_{\mu_1\cdots\mu_p)} \,.
\ee
Since $f_\mu$ must be given uniquely, $f_\mu$ is written by $f_\mu=\partial_\mu \ln \psi$.
Hence, if $K_{\mu_1\cdots\mu_p}$ is constructible, then $f_\mu$ is closed.\hfill$\Box$\\

Given a rational first integral, we obtain a pair of generalised Killing tensors with the common associated 1-form.
Proposition II.3 states that
by investigating whether the associated 1-form
is closed or not, we can check whether
a rational frist integral is inconstructible.
Using this fact, we investigate
several concrete examples of rational first integrals
in the next two sections.

\subsection{Geometric formulation of generalised Killing tensors}
Let us introduce the connection ${\cal D}_\mu$ on $\bigotimes^n T^*M$
which acts on a tensor $T_{\mu_1\dots\mu_n}$ as
\be
{\cal D}_\mu T_{\nu_1\dots\nu_n} = \nabla_\mu T_{\nu_1\dots\nu_n}
- \sum_{i=1}^n A_{(\mu} 
T_{|\nu_1\cdots\nu_{i-1}|\nu_i)\nu_{i+1}\dots\nu_n} \,,
\label{connection_2}
\ee
where $A_\mu$ is a 1-form.
This connection is torsion-free, and
the metric condition does not hold, ${\cal D}_\mu g_{\nu\rho}\neq 0$.
Hence, the curvature tensor of ${\cal D}_\mu$, defined by
${\cal R}_{\mu\nu\rho}{}^\sigma V_\sigma\equiv
({\cal D}_\mu {\cal D}_\nu-{\cal D}_\nu {\cal D}_\mu) V_\rho$,
has antisymmetry with respect to the initial two indices,
${\cal R}_{\mu\nu\rho}{}^\sigma=-{\cal R}_{\nu\mu\rho}{}^\sigma$,
and the Bianchi identities ${\cal R}_{[\mu\nu\rho]}{}^\sigma=0$
while antisymmetry of the latter two indices does not hold.
Using this connection, we can show that
the generalised Killing tensor equation \eref{GKSeq2}
is written in the form
\be
{\cal D}_{(\mu}K_{\nu_1\dots \nu_p)} = 0 \,, \label{GKSeq3}
\ee
with the identification of $f_\mu=p A_\mu$.
The point is that this equation has same form as the ordinary Killing tensor equation,
where the Levi-Civita connection $\nabla_\mu$ is replaced
by the present connection ${\cal D}_\mu$.
This fact is suggestive to generalise other hidden symmetries.

Since ordinary Killing tensors form a (graded) Lie algebra
with respect to the Schouten-Nijenhuis (SN) bracket of the Levi-Civita connection
\cite{Dubois-Violette:1995,Cariglia:2011},
it would be interesting to ask whether generalised Killing tensors do as well.
Unfortunately, it fails with respect to the SN bracket of neither the Levi-Civita connection
nor the present connection.

\subsection{Integrability conditions}
An application of introducing the torsion-free connection \eref{connection_2} is that
the integrability conditions for generalised Killing tensors
can be written in a simple form.
For simplicity, let us consider generalised Killing vectors,
which are given by
\be
{\cal D}_{(\mu}\xi_{\nu)}= 0 \,. \label{GKV}
\ee
The integrability conditions for generalised Killing vectors
were already provided by Collinson\cite{Collinson:1986},
which are written in terms of the Riemann tensor and the associated 1-form
but the expressions provided are rather complicated.
On the other hand, we now obtain from Eq.\ \eref{GKV} the equations
\ba
{\cal D}_\mu \xi_\nu &=& L_{\mu\nu} \,, \\
{\cal D}_\mu L_{\nu\rho} &=&
- {\cal R}_{\nu\rho\mu}{}^\sigma \xi_\sigma \,,
\ea
where $L_{\mu\nu}\equiv {\cal D}_{[\mu}\xi_{\nu]}$.
Hence, the integrability conditions
for generalised Killing vectors are given by
\be
{\cal D}_{[\mu}{\cal R}_{|\rho\sigma|\nu]}{}^\lambda \xi_\lambda
+ {\cal R}_{\mu\nu[\rho}{}^\lambda L_{\sigma]\lambda}
+ {\cal R}_{\rho\sigma[\mu}{}^\lambda L_{\nu]\lambda} = 0 \,,
\ee
which has same form as those for ordinary Killing vectors
but the Riemann tensor $R_{\mu\nu\rho}{}^\sigma$ has been
replaced by the curvature tensor ${\cal R}_{\mu\nu\rho}{^\sigma}$.
This is also available for generalised Killing tensors of arbitrary rank.
The integrability conditions
for second-rank Killing tensors have been provided
in terms of the Riemann tensor \cite{Hauser:1975}.
Replacing the Riemann tensor with
the curvature tensor of ${\cal D}_\mu$,
we will obtain the integrability conditions
for second-rank generalised Killing tensors.

\section{Collinson-O'Donnell solution}
Vaz and Collinson \cite{Vaz:1992} have found the canonical forms for the metrics of spacetimes
in four dimensions admitting a pair of generalised Killing vectors
under the assumption that one of the generalised Killing vectors is hypersurface orthogonal.
Using one of the results,
which is the case when one of the generalised Killing vectors is null
and not orthogonal to another, Collinson and O'Donnell \cite{Collinson:1992}
have obtained the solutions of the vacuum Einstein equations,
which were classified into two cases.
The solution of Case 2 was given in the form
\ba
ds^2 &=& - \frac{y}{x}dtdx + \frac{yt}{x^2}dx^2
+ \frac{\alpha^2}{2\sqrt{y}}(dy^2+dz^2) \\
& & - \frac{2\sqrt{y}\,\alpha^2}{x}dx\left(
\frac{f}{\sqrt{y}} dy - g \sqrt{y} dz\right) \,, \nonumber
 \label{metric_CD_case2}
\ea
where $\alpha$ is a constant,
$f$ and $g$ are arbitrary functions of $y$ and $z$ satisfying
the differential equations
\be
\partial_yf-y \partial_zg = -\frac{C^2}{y^2\sqrt{y}} \,, \quad
\partial_zf+y \partial_yg = \frac{C}{y\sqrt{y}} \label{cond_fg}
\ee
with a constant $C$.
For this metric, the geodesic equations
admit an irreducible rational first integral
\be
F = \frac{p_x}{p_t} \,,
\ee
and a pair of generalised Killing vectors
$\partial_t$ and $\partial_x$ can be found
with the common associated 1-form $(2/x) dx$.
We note that $F$ is a rational first integral
even if $f$ and $g$ do not satisfy Eq.\ \eref{cond_fg}.
Since the associated 1-from is closed, 
we find from Proposition II.3 that
the rational first integral is constructible.
Indeed, we find that
$x \partial_t$ and $x \partial_x$ are independent Killing vectors,
and the rational first integral is given by $F=Q_2/Q_1$
with two independent polynomial first integrals $Q_1=x p_t$ and $Q_2=x p_x$.

The solution of Case 1 is obtained as the limiting case.
Indeed, if we take $y\to 1 + \epsilon y$ and $z\to\epsilon z$
with $f\to \epsilon f$, $g\to\epsilon g$ and
$\alpha^2\to\alpha^2/\epsilon^2$ and then send $\epsilon\to 0$,
we obtain the metric
\ba
ds^2 &=& - \frac{1}{x}dtdx + \frac{t}{x^2}dx^2 \nonumber\\
& &+ \frac{\alpha^2}{2}(dy^2+dz^2)
 - \frac{2\alpha^2}{x}dx(
      f dy-g dz) \,, \label{metric_CD_case1}
\ea
where $f$ and $g$ are functions of $y$ and $z$ satisfying
\be
\partial_yf- \partial_zg = -C^2 \,, \quad
\partial_zf+ \partial_yg = C \,.
\ee
Since this metric still has two independent Killing vectors $x \partial_x$ and $x \partial_t$, two generalised Killing vectors
$\partial_t$ and $\partial_x$ are not inconstructible.
It consequently follows that the rational first integral is not inconstructible.

\section{Metrics admitting an inconstructible rational first integral}
\subsection{Two dimensions}
To construct metrics admitting an inconstructible rational first integral in two dimensions,
we consider the Maciejewski-Przybylska system \cite{Maciejewski:2004}.
The Hamiltonian is given by
\be
 H = \frac{1}{2}(p_x^2+p_y^2)
     + f(p_x,p_y) (x p_x-\alpha yp_y) \,, \label{Hamiltonian_MP}
\ee
where $\alpha$ is a constant and $f(p_x,p_y)$ is a function of $p_x$ and $p_y$.
The Hamiltonian admits a first integral of the form 
\be
 F = p_x^\alpha p_y \label{FI_MP}
\ee
for arbitrary $\alpha$ and $f$.
To obtain a rational first integral,
we assume that $\alpha$ is a negative rational number.
Actually, setting $\alpha=-s/r$ with relatively prime,
positive integers $r$ and $s$,
we have a rational first integral $F^r = p_y^r/p_x^s$.
Moreover, we take $f=p_x+p_y$ to make the Hamiltonian quadratic.
Then, the Hamiltonian describes geodesic flows on a two-dimensional surface
with the metric
\be
 ds^2 = \frac{(1-2 \alpha y)dx^2 
 	-2 (x- \alpha y)dx dy+(1+2x)dy^2 }{Q(x,y)} \,, \label{metric1}
\ee
where
\be
 Q(x,y) = 1+2x-2 \alpha y-(x+\alpha y)^2 \,. \label{func_Q}
\ee
A single parameter $\alpha$ is contained.

First, we focus on the case when $\alpha=-1$.
In this case, the metric \eref{metric1} is flat.
Since the first integral is given by $F=p_y/p_x$,
$\partial_x$ and $\partial_y$ are generalised Killing vectors.
The common associated 1-form
is dual to $- (\partial_x + \partial_y)$.
Since the associated 1-form is closed,
it turns out from Proposition 2.3 that $\partial_x$
and $\partial_y$ are constructible.
More explicitly, we perform the coordinate transformation
\be
 x=u+v+\frac{1}{2}(u^2+v^2) \,, \quad
 y=u-v+\frac{1}{2}(u^2+v^2) \,.
\ee
In the $(u,v)$ coordinates, the metric is given by $ds^2=du^2+dv^2$
and the generalised Killing vectors are given by
\ba
&& \partial_x = \frac{1-v}{1+u}\partial_u + \partial_v
         = \frac{1}{1+u}\left(
           \partial_u +\partial_v - (v\partial_u-u\partial_v)\right) \,, \\
&& \partial_y = \frac{1+v}{1+u}\partial_u - \partial_v
         = \frac{1}{1+u}\left(
           \partial_u-\partial_v +(v\partial_u-u\partial_v) \right) \,.
\ea
Indeed, they are constructible.

For general $\alpha=-s/r$, since $F^r=p_y^r/p_x^s$ is a rational first integral,
$(\partial_x)^s$ and $(\partial_y)^r$
are respectively generalised Killing tensors
with the common associated 1-form which is dual to $-s (\partial_x +\partial_y)$.
Since the associated 1-form is not closed except for $\alpha=-1$,
the rational first integral is inconstructible for $\alpha\neq -1$.
Thus, we have constructed the metric \eref{metric1}
admitting an inconstructible rational first integral of the geodesic equations in two dimensions,
with the exception for the flat case ($\alpha=-1$).

\subsection{Four dimensions}
We are able to generalise the Maciejewski-Przybylska system \eref{Hamiltonian_MP}
to the $n$-dimensional system.
The Hamiltonian is given by
\be
 H = \frac{1}{2}\sum_{i=1}^n p_i^2 
 + f(p_1,\dots,p_n) \sum_{i=1}^n \alpha_i x^i p_i \,, \label{Hamiltonian_MP_Ddim}
\ee
where $\alpha_1,\dots,\alpha_n$ are constants and
$f(p_1,\dots,p_n)$ is a function of $p_1,\dots,p_n$.
The Hamiltonian admits a first integral
\be
 F = p_1^{\beta_1}p_2^{\beta_2}\cdots p_n^{\beta_n} \,, \label{rfi_MP}
\ee
where $\beta_1,\dots,\beta_n$ are constants satisfying the condition
\be
 \alpha_1\beta_1 + \alpha_2\beta_2 + \dots + \alpha_n\beta_n = 0 \,.
\ee
We remark that this system is integrable because Eq.\ \eref{rfi_MP} describes $n-1$ constants of motion at the same time.
In fact, introducing $n-1$ constants $Q_i$ $(i=2,\dots,n)$,
we obtain the relations
$p_i = Q_i p_1^{\alpha_i/\alpha_1}$ for $i=2,\dots,n$.
Moreover, substitute these relations into the Hamiltonian
and consider the energy $H=Q_1$, then, in principle,
all the momenta can be written as functions of $x^\mu$
including $n$ constants $Q_i$, $p_i = p_i(x^1,\dots,x^n;Q_1,\dots,Q_n)$.

When we take $f$ as $f=a_1 p_1+a_2p_2 + \dots + a_n p_n$,
where $a_1,\dots,a_n$ are constants,
the Hamiltonian \eref{Hamiltonian_MP_Ddim} describes
geodesic flows on the $n$-dimensional curved space
with the inverse metric
\be
 g^{ii} = 1 + 2 a_i\alpha_i x^i \,, \quad
 g^{ij} = a_j \alpha_i x^i + a_i\alpha_j x^j \,.
\ee
For simplicity, let us consider the Maciejewski-Przybylska
in four dimensions.
Moreover, we adopt the following setup:
$a_1=a_2=a_3=1$, $a_4=-\sqrt{-1}$, $\alpha_1=1$, $\alpha_2=-\alpha$ and
$\alpha_3=\alpha_4=0$.
Under this setup, the Hamiltonian
is independent of the coordinates $x^3$ and $x^4$,
so that $p_3$ and $p_4$ are first integrals.
Since another first integral is given by $F=p_1^{\beta_1}p_2^{\beta_2}$
with $\beta_1-\beta_2 \alpha=0$,
we normalise $\beta_2$ as $\beta_2=1$ and then obtain $\beta_1=\alpha$.
Moreover, identifying the coordinates $x^1,x^2,x^3$ as $x,y,z$ and $x^4$
as $\sqrt{-1}\,w$, we obtain the Hamiltonian
\ba
 H &=& \frac{1}{2}(p_x^2+p_y^2+p_z^2-p_w^2) \nonumber\\
   & &  + (p_x+p_y+p_z+p_w)(xp_x-\alpha yp_y)
\ea
with the first integrals $p_z$, $p_w$ and $F=p_x/p_y^\alpha$.
Hence, we obtain one-parameter family of four-dimensional metrics
 admitting integrable geodesic flows.
In particular, when we take $\alpha=-1$, the metric becomes scalar-flat.
Namely, the scalar curvature vanishes while the Ricci tensor is nonzero.
The components of the scalar-flat metric are given by
\ba
&& g_{xx} = \frac{1+2y}{K(x,y)} \,, \quad
   g_{yy} = \frac{1+2x}{K(x,y)} \,, \quad
   g_{xy} = \frac{-x-y}{K(x,y)} \,, \nonumber\\
&& g_{zz} = \frac{1+2x+2y+2xy}{K(x,y)} \,, \quad
   g_{zw} = \frac{-x^2-y^2}{K(x,y)} \,, \nonumber\\
&& g_{xz} = \frac{y^2-xy-x}{K(x,y)} \,, \quad
   g_{yz} = \frac{x^2-xy-y}{K(x,y)} \,, \\
&& g_{ww} = \frac{2(x-y)^2-1-2x-2y+2xy}{K(x,y)} \,, \nonumber\\
&& g_{xw} = \frac{-y^2+xy+x}{K(x,y)} \,, \quad
g_{yw} = \frac{-x^2+xy+y}{K(x,y)} \,, \nonumber \label{metric2}
\ea
where
\be
 K(x,y) = 1+2x+2y+2xy-x^2-y^2 \,.
\ee
This metric admits a rational first integral $F=p_y/p_x$,
which is inconstructible.
Thus, we have constructed the scalar-flat metric \eref{metric2}
admitting an inconstructible rational first integral of the geodesic equations in four dimensions.
We expected the scalar-flat metric being a solution
of the Einstein-Maxwell equations in four dimensions,
but it ended up in failure unfortunately.

\section{Generalised hidden symmetries}

\subsection{Generalised conformal Killing tensors}
For the integration of the equations of motion in a constrained system with $H=0$,
it is sufficient to find a quantity conserved at least along the zero energy orbits.
Denoted by $F$, such a quantity is expressed by the condition $\{H,F\}=LH$ with some function $L$.
If we consider $F$ as polynomial in momenta,
the condition leads to the conformal Killing tensor equation
\be
\nabla_{(\mu}K_{\nu_1\dots \nu_p)}
= g_{(\mu\nu_1}N_{\nu_2\cdots\nu_p)} \,, \label{CKS}
\ee
where $N_{\mu_1\cdots\mu_{p-1}}$ is a symmetric tensor.
In analogy to this, we consider $F$ as a rational quantity in momenta.
When $F$ is given by Eq.\ \eref{basic1}, the condition is written as
\be
 \{H,P\}=AP+L_1H \,, \quad
 \{H,Q\}=AQ+L_2H \,,
\ee
where $A$, $L_1$ and $L_2$ are some functions
related to $L$ by $L=L_1Q-L_2P$.
Hence, writing $P$, $A$ and $L_1$ as
$P=K^{\mu_1\dots\mu_p}p_{\mu_1}\dots p_{\mu_p}$,
$A=f^\mu p_\mu$ and $L_1=N^{\mu_1\dots\mu_{p-1}}p_{\mu_1}\dots p_{\mu_{p-1}}$,
we obtain the generalised Killing tensor equation
\be
 {\cal D}_{(\mu}K_{\nu_1\dots \nu_p)}
 = g_{(\mu\nu_1}N_{\nu_2\cdots\nu_p)} \,, \label{GCKS}
\ee
where ${\cal D}_\mu$ is the connection given by Eq.\ \eref{connection_2}.
If ${\cal D}_\mu$ were the Levi-Civita connection,
this equation would be the conformal Killing tensor equation.
Hence, we call the symmetric tensor $K_{\mu_1\dots\mu_p}$ satisfying
the equation \eref{GCKS} a generalised conformal Killing tensor.
In the same manner as generalised Killing tensors, we need to find a pair
of generalised conformal Killing tensors with a common connection
to obtain a rational quantity conserved only along the zero energy orbits.

\subsection{Generalised (conformal) Killing-Yano tensors}
Employing the connection ${\cal D}_\mu$,
we may define the generalised Killing-Yano tensor
$f_{\mu_1\cdots\mu_p}$ by the differential equation
\be
 {\cal D}_{(\mu}f_{\nu_1)\nu_2\cdots\nu_p} = 0 \,, \label{GKY}
\ee
where $f_{\mu_1\cdots\mu_p}$ is a $p$-form.
If the connection ${\cal D}_\mu$ is the Levi-Civita connection,
this equation is the Killing-Yano equation.
In the similar fashion, we are also able to define
the generalised conformal Killing-Yano tensor $f_{\mu_1\cdots\mu_p}$
by the differential equation
\be
 {\cal D}_{(\mu}f_{\nu)\rho_1\cdots\nu_{p-1}}
= g_{\mu\nu}\xi_{\rho_1\cdots \rho_{p-1}}
  +\sum_{i=1}^{n-1} (-1)^i g_{\rho_i(\mu}
\xi_{\nu)\rho_1\cdots\hat{\rho}_i\cdots\rho_{p-1}}  \,, \label{GCKY}
\ee
where the hat ($\hat{~}$) eliminates the index and
\be
 \xi_{\mu_1\cdots\mu_{p-1}}
= \frac{1}{n-p+1} {\cal D}^\nu f_{\nu\mu_1\cdots \mu_{p-1}}
\ee
is called the associated $(p-1)$-form of $f_{\mu_1\cdots\mu_p}$
and $n$ is the dimension of a space or spacetime.
If the connection ${\cal D}_\mu$ is the Levi-Civita connection,
this equation is the conformal Killing-Yano equation.
It is remarkable that these generalised (conformal) Killing-Yano tensors
are also related to rational first integrals of geodesic equations
because the "square" of a generalised (conformal) Killing-Yano
tensor, $K_{\mu\nu}\equiv f_{\mu\rho_1\cdots \rho_{p-1}}
f_\nu{}^{\rho_1\cdots\rho_{p-1}}$, becomes
a generalised (conformal) Killing tensor.
Further investigation are left as a future problem.

\section{Summary and discussion}
In this paper, we have discussed rational first integrals of geodesic equations.
We introduced the notion of inconstructible rational first integrals,
which cannot be constructed from two polynomial first integrals,
and showed in Proposition 2.3 that a rational first integral is
not inconstructible if and only if the associated 1-form of the generalised Killing tensors
read from the rational first integral is closed.
Using this fact, we showed that the rational first integral of the Collinson-O'Donnell solution is not inconstructible.
We also constructed several examples for metrics in two and four dimensions
admitting an inconstructible first integral of geodesic equations
by using the Maciejewski-Przybylska system.
In particular, we obtained a scalar-flat metric in four dimensions.
Unfortunately, the scalar-flat metric is not a solution of the Einstein-Maxwell equations.
Hence, it would be an important task in general relativity
to construct a physically interesting solution of Einstein equations.

We have discussed novel generalisations of hidden spacetime symmetries,
which are related to rational first integrals of geodesic equations.
The generalised Killing tensors \eref{GKSeq3} are defined
by the Killing equation with the Levi-Civita connection
replaced by the torsion-free connection \eref{connection_2}.
In the similar fashion, we introduced the generalised
conformal Killing tensors \eref{GCKS}, Killing-Yano tensors \eref{GKY}
and conformal Killing-Yano tensors \eref{GCKY}.
In this paper, we have worked based on some geometric aspects of hidden symmetries.
However, as the concept of hidden spacetime symmetries have helped us
in understanding various gravitational phenomena especially in black hole physics,
it would be meaningful to consider their applications.

\section*{Acknowledgements}
The authors would like to thank Jiro Soda for useful comments. This work
was supported by the JSPS Grant-in-Aid for Scientific Research No.\ 26$\cdot$1237.

\end{document}